# A Transferable Recommender Approach for Selecting the Best Density Functional Approximations in Chemical Discovery

Chenru Duan[1,2], Aditya Nandy[1,2], Ralf Meyer[1], Naveen Arunachalam[1], and Heather J. Kulik[1,2]

[1]Department of Chemical Engineering, Massachusetts Institute of Technology, Cambridge, MA 02139

[2]Department of Chemistry, Massachusetts Institute of Technology, Cambridge, MA 02139

**Abstract:** Approximate density functional theory (DFT) has become indispensable owing to its cost-accuracy trade-off in comparison to more computationally demanding but accurate correlated wavefunction theory. To date, however, no single density functional approximation (DFA) with universal accuracy has been identified, leading to uncertainty in the quality of data generated from DFT. With electron density fitting and transfer learning, we build a DFA recommender that selects the DFA with the lowest expected error with respect to gold standard but cost-prohibitive coupled cluster theory in a system-specific manner. We demonstrate this recommender approach on vertical spin-splitting energy evaluation for challenging transition metal complexes. Our recommender predicts top-performing DFAs and yields excellent accuracy (ca. 2 kcal/mol) for chemical discovery, outperforming both individual transfer learning models and the single best functional in a set of 48 DFAs. We demonstrate the transferability of the DFA recommender to experimentally synthesized compounds with distinct chemistry.

**Introduction**

With rapid advancements in computing power, density functional theory (DFT) has become an indispensable companion to experiments as well as a primary tool in virtual high throughput screening (VHTS)[1] to generate large-scale computational datasets[2]. Combined with machine learning (ML), these datasets have accelerated computational chemical discovery and revolutionized scientific discoveries[3, 4] in physics, chemistry, materials sciences, and biology. All ML models, however, are limited by the quality of the training data, imposing strict requirements for the accuracy of the first-principles methods used in VHTS. In DFT, a density functional approximation (DFA) that works well on certain systems can fail prominently on other systems due to the approximations made in the exchange-correlation functional[5]. This DFA dependence is particularly strong in open-shell transition metal chemistry, with many examples of compelling functional materials (e.g., metal organic frameworks) and catalytic reactions (e.g., C–H activation) that are dominated by static correlation[6].

For simplicity, a single DFA is typically selected to screen through a large chemical space in VHTS[2]. It is known that the single-DFA approach can lead to bias in the computational data sets generated, which further bias ML models and the final candidate materials in the ML-accelerated discovery[7]. Alternatively, when computational chemistry efforts are focused enough on a small set of molecules for which experimental or accurate correlated wavefunction theory reference data is available, different DFAs may be evaluated and then selected in a system and property-dependent manner to obtain agreement[8]. Nevertheless, there is no guarantee that the same DFA is optimal for different properties or different materials. These challenges pose limitations on both mechanistic inquiry by computational chemists aiming to reach the "chemical accuracy" needed to be predictive of experimental outcomes and in the emergence of massive-scale "big data" that enables deep model training in computational sciences[9] as in computer vision and natural language processing.

One way to improve the fidelity of DFT-derived data sets is to develop DFAs with increased accuracy and generalizability. Recent advances in using ML to develop neural network potentials[10] and exchange-correlation functionals[11] have shown promise in developing transferable models. Currently, however, they have severe limitations that curtail their practical use. These DFAs have primarily been developed for and applied on narrow sets of closed-shell organic molecules and are still less transferable relative to conventional DFAs developed in the theoretical chemistry community over the past few decades. More importantly, these models only target the total electronic energy of a geometry rather than other properties of chemical interest, such as those involving multiple electronic states (e.g., spin splitting).

Here, we demonstrate an alternate route. Instead of developing a new DFA, we leverage transfer learning (TL) and use a "regress-then-classify" strategy to develop a recommender[12] to select the best conventional DFA for a given system and a property of interest based on features the electron density of the system under study. To demonstrate this approach, we recommend a DFA that most accurately evaluates the vertical spin splitting energy of a transition metal complex (TMC), a property highly sensitive to the choice of DFA[7,20]. This recommender selects a DFA in a system-specific manner rather than based on average performance of DFAs and captures the rank ordering of the top-performing DFAs. Crucially, our recommender has a mean absolute error (MAE) of 2.1 kcal/mol, providing the accuracy required for the exploration of transition metal chemical space. Because the electron density is a fundamental property of the system, our recommender demonstrates excellent transferability, achieving similar accuracy for recommending DFAs on unseen experimentally synthesized TMCs that contain diverse chemistry. This recommender approach is expected to be a general framework for method selection to increase the data quality in VHTS and ML-accelerated discovery in computational sciences.

**Results**

**Overview of the DFA recommender.** Electron density lies in the core of Kohn-Sham (KS)-DFT and can derive any ground state property of interest for a system. However, it is less commonly used as representations in ML models due to its non-local nature in KS orbitals and its cumbersome to fulfill translational and rotational symmetries when discretized to 3D cubes. Here, we perform density fitting on the electron density[13] of a TMC obtained from the representative B3LYP calculation. The resulting coefficients of the basis functions, which preserve the required physical (i.e., translational, rotational, and permutation) symmetries, are used as the representations of the TMC (Fig. 1a–1b, see *Methods*). To make this approach more general with respect to chemical properties that involve multiple electronic states (here, vertical spin splitting), we directly decompose the difference of electron density at two electronic states. These density fitting coefficients are atom-centered, which have the advantage of being agnostic to model architectures and can be readily combined with Behler–Parrinello, message passing, and graph neural networks.

With the overarching goal of recommending a DFA in a system-specific way, one may expect that the optimal approach is to treat the learning task as a multi-class classification problem[14] where the classes are different DFAs. However, due to the inherent similarity among different DFAs, many DFAs perform similarly well in most cases, leading to label noise when determining the "best" DFA. Correspondingly, this noise poses a challenge to classification models (Supplementary Fig.1). Therefore, we take an alternative "regress-then-classify" strategy, where we aim to recommend a low-error DFA instead of forcing ML models to recognize top-ranking DFAs (Fig. 1, see *Methods*). We thus set up transfer learning (TL) regression task to predict the absolute difference of the vertical spin splitting energy between a DFA (*f*) and our reference data from domain-based local pair natural orbital (DLPNO)-CCSD(T) theory[15] ($|\Delta\Delta E_{H-L}[f]|$). In this "regress-then-classify" strategy, we first build Behler–Parrinello-type neural networks to predict $|\Delta\Delta E_{H-L}[f]|$ using our B3LYP density fitting coefficients separately for a pool of pre-selected candidate DFAs (Fig. 1c). We then select the DFA that gives the smallest absolute predicted difference (i.e., predicted $|\Delta\Delta E_{H-L}[f]|$) as the recommend DFA for the TMC (Fig. 1d).

There are two approaches to evaluate the performance of the DFA recommender. The first is the absolute error as a result of using the recommended DFA relative to the reference DLPNO-CCSD(T) method. This measure serves as a practical metric for evaluating the accuracy obtained by the recommender in VHTS. The second is the rank ordering of the recommended DFA among the pool of candidate DFAs. This statistical measure quantifies how well the recommender distinguishes top-performing candidate DFAs. Throughout our work, we will consider both perspectives.

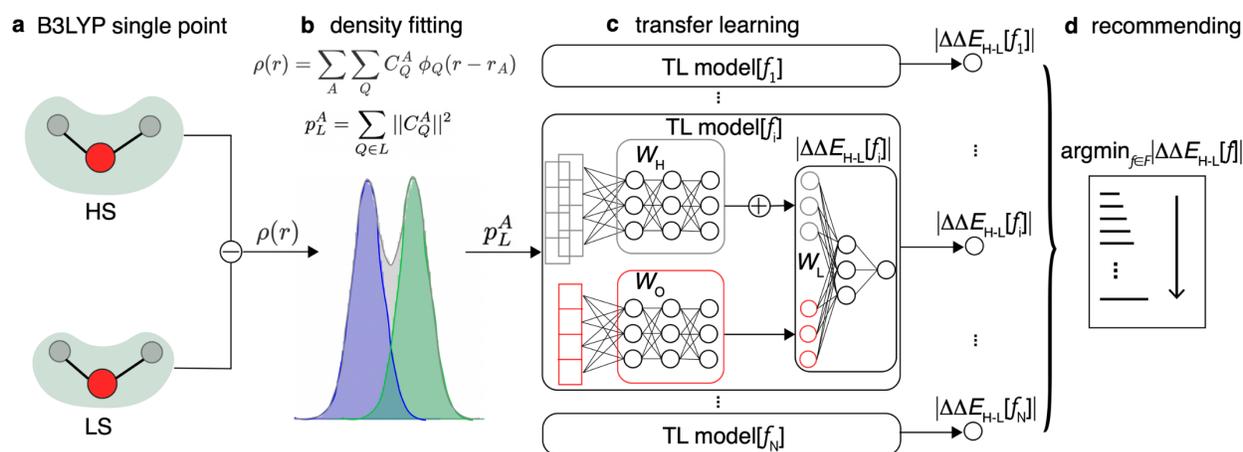

**Fig. 1 | Workflow for the DFA recommender. a**, B3LYP/def2-TZVP single-point energy calculations are performed on both the high-spin (HS) and low-spin (LS) states to obtain their electron densities at B3LYP level. **b**, The difference of the electron densities between the HS and LS states is decomposed into each atom using a density fitting procedure. **c**, These coefficients are used in a Behler–Parrinello-type neural network as a TL model to predict $|\Delta\Delta E_{H\text{-}L}[f]|$ for each DFA ($f$) in our pool of 48 DFAs. Coefficients of different atoms that are in the same group of the periodic table share the same local network and weights (e.g., $W_O$ for O (red), $W_H$ for H (gray) and O and S share the same weights $W_O$). The latent vectors of each element are lastly concatenated and passed to a fully-connected network ($W_L$) to predict $|\Delta\Delta E_{H\text{-}L}[f]|$. **d**, The 48 predicted $|\Delta\Delta E_{H\text{-}L}[f]|$ are then sorted, where we recommend the DFA that yields the lowest predicted $|\Delta\Delta E_{H\text{-}L}[f]|$.

**Performance of TL models.** We first consider a set of 452 octahedral TMCs composed of $3d$ mid-row transition metals and small common organic ligands in spectrochemical series (*VSS-542*, Supplementary Fig. 2, see *Methods*). We demonstrate the performance of our TL models for predicting the differences in vertical spin-splitting energies obtained by a DFA and DLPNO-CCSD(T) for a pool of 48 DFAs that cover multiple rungs of "Jacob's ladder". These 48 TL models have mean absolute errors (MAEs) ranging from 2.3 kcal/mol to 3.4 kcal/mol, with a median MAE of 2.5 kcal/mol on the 152 set-aside test TMCs in *VSS-452* (Fig. 2a). These MAEs are low considering the fact that the TL models were only trained on a small set of 300 TMCs that contain diverse chemistry (see *Methods*). In fact, these MAEs are lower than the typical experimental uncertainties for thermochemical properties of TMCs (e.g., 3 kcal/mol)[16]. Despite the fact that some DFAs have very large (i.e., > 30 kcal/mol) DFA-derived MAEs, all 48 TL models yield reasonably low MAEs, suggesting the general applicability of this TL approach regardless of the candidate DFA considered.

Interestingly, the ranking of TL model MAEs does not have the same order as the error ranking of the underlying DFA results relative to DLPNO-CCSD(T). For example, the DFA with the lowest TL MAE is a double-hybrid functional DSD-PBEB95-D3BJ, which only has the fifth-lowest MAE relative to the reference calculation (Supplementary Figs. 3 and 4). MN15, which gives the highest MAE among 48 TL models, only ranks 17th among the DFA MAEs relative to DLPNO-CCSD(T). For the set of 48 DFAs, the rank-order coefficient (i.e., Spearman's $r$) between DFAs ranked by TL model MAE and those ranked by DFA-derived MAE is only 0.36. This observation suggests that a TL model does not necessarily have improved performance when the MAE of the DFA-derived MAE of the baseline DFA is smaller, posing an interesting question of how to select the best baseline method from which TL would yield the lowest errors.

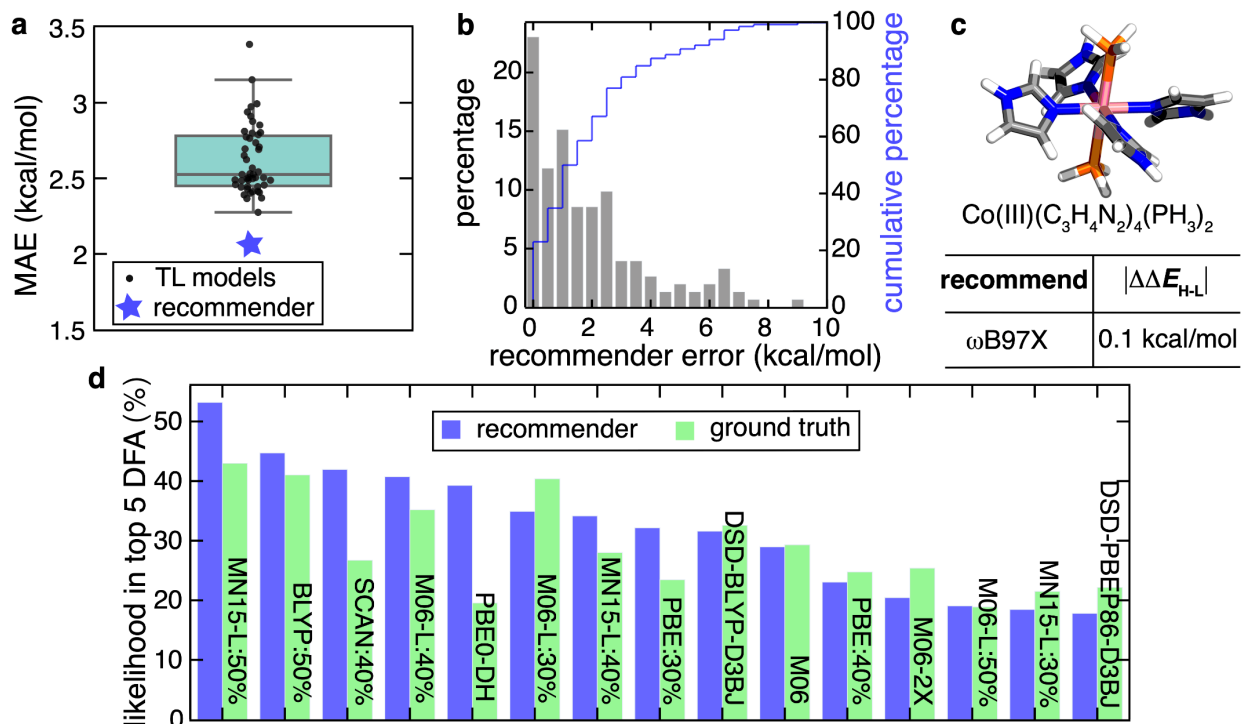

**Fig. 2 | Performance of TL models and the recommender on *VSS-452* set. a**, MAE of 48 TL

models for the prediction of $|\Delta\Delta E_{\text{H-L}}|$ with a box indicating their median (solid line). The MAEs for each TL model (black circle) and our recommender approach (blue star) are also shown. **b**, Percentage for the absolute error with the recommender-selected DFA (gray bars), with the cumulative percentage (blue solid line) shown according to the axis on the right. **c**, An example complex Co(III)(C$_3$H$_4$N$_2$)$_4$(PH$_3$)$_2$, its recommended DFA (i.e ωB97X), and the associated DFA error (0.1 kcal/mol). Atoms are colored as follows: pink for Co, brown for P, gray for C, blue for N, and white for H. **d**, Percentage likelihood of a DFA residing in the top-5 choices suggested by ground truth (green) and our DFA recommender (blue). The DFAs are sorted in a descending order of the predicted likelihood of the recommender. In all cases, the model performance is evaluated on the set-aside 152 test complexes of *VSS-452*.

**Performance of the DFA recommender.** We then utilize the predicted $|\Delta\Delta E_{\text{H-L}}[f]|$ from all 48 TL models to recommend a "best" DFA. We compare the vertical spin splitting energy obtained by the recommended DFA and the DLPNO-CCSD(T) reference to evaluate the performance of the recommender (see *Methods*). The recommender achieves an MAE of 2.1 kcal/mol on the set-aside test set of *VSS-452*, outperforming all 48 TL models. This MAE is only 2.5 times the theoretical lower bound (0.8 kcal/mol), which is the performance we would achieve if the DFA that gives the lowest error is always selected (Supplementary Fig. 5). Our recommender outperforms random DFA selection (MAE = 13.3 kcal/mol) by a factor of 6.5. More importantly, even for an alternate strategy representative of using prior knowledge in which we pick the single DFA with the lowest average error over the *VSS-452* set, its DFA-derived MAE (i.e., 6.2 kcal/mol with DSD-BLYP-D3BJ) would be three times larger than that of our recommender. With this recommender, we are able to select DFAs that give errors within the 3 kcal/mol threshold required for transition metal chemical discovery in 77% cases (Fig. 2b). Almost for all complexes (i.e., 94%), the recommender can achieve a higher accuracy than the MAE (i.e., 6.2 kcal/mol) of the best single DFA.

One distinct advantage of this recommender approach is that its performance is likely to improve systematically with increasing number of DFAs under consideration, despite using the same training data set and TL models. For example, the recommender would achieve an MAE of 3.0 kcal/mol if we had used the smaller pool of 23 candidate DFAs introduced in our previous work[7] (Supplementary Table 1). However, as we add the remaining 25 DFAs that incorporate alternate fractions of Hartree-Fock (HF) exchange, the MAE reduces to 2.1 kcal/mol, improving upon the accuracy of our best TL model of DSD-PBEB95-D3BJ (2.3 kcal/mol, Supplementary Table 2). We gain this additional accuracy of the recommender without significantly increasing the computational cost, as there is no need for more training data with the computationally demanding DLPNO-CCSD(T) reference.

The other distinct feature of our recommender approach is its system specificity. Compared to the widely used strategy of selecting DFAs based on the statistically averaged performance over a benchmark data set[17], the recommender chooses the DFA only based on the chemistry (here the electron density) of the system under consideration. As a result, our recommender can avoid selecting a DFA that performs well on average yet particularly bad on the given complex. For example, DSD-BLYP-D3BJ has the lowest DFA-derived MAE for $|\Delta\Delta E_{H-L}[f]|$ (i.e., 6.2 kcal/mol) against DLPNO-CCSD(T) but gives a relatively high absolute error of 9.2 kcal/mol on $Co(III)(C_3H_5N_2)_4(PH_3)_2$. Our recommender, instead, selects ωB97X, which has a very small error of 0.1 kcal/mol on this compound, despite the fact the DFA-derived MAE of ωB97X is much higher (i.e, 16.9 kcal/mol) and only ranks 37[th] out of 48 DFAs (Fig. 2c and Supplementary Fig. 3).

Next, we investigate the statistics of the recommended DFA performance to determine when our recommender correctly selects the top-performing functional. We focus on DFAs that are within the top-5 choices as they usually result in the accuracy required for studies in transition metal chemistry[16] (i.e., 3.0 kcal/mol) because multiple DFAs can achieve similar accuracy for a TMC (Supplementary Figs. 5 and 6). We find more than two-thirds (i.e., 100 out of 152 complexes in the set-aside test set of *VSS-452*) of the recommended DFAs are within the top-5 DFAs relative to the ground truth (Supplementary Fig. 7). In less than 15% of cases, our approach recommends a DFA that is not in the top 10 out of 48 candidate DFAs. Interestingly, we get more favorable ranking statistics using only the 23 DFAs from prior work[7], despite a higher recommender MAE (i.e., 3.0 kcal/mol, Supplementary Table 1). With these 23 candidate DFAs, 88% of the DFAs selected by the recommender are within the top 5 and nearly none (i.e. 4%) fall out of the top-10 DFAs (Supplementary Fig. 8). We expect this behavior to be general in our recommender approach: with more candidate DFAs in the pool, it is more difficult to get favorable ranking statistics, as judged by identifying the single top performing functional, but easier to obtain a lower MAE for practical performance because there are more DFAs to select from.

Lastly, we compare the statistics of most probable DFAs that reside in top-5 choices obtained by the ground truth and our recommender. Out of the 48 DFAs, MN15-L with 50% HF exchange (i.e., MN15-L:50%) appears most frequently as a top-5 DFA, with a likelihood of 43% for the 152 set-aside test TMCs in *VSS-452* (Fig. 2d). Correspondingly, our recommender identifies the same DFA to have the highest likelihood (53%) being in the top-5 candidate DFAs. Moreover, for each base semi-local functional (i.e., BLYP, PBE, SCAN, M06-L, and MN15-L), the DFA recommender successfully picks the HF exchange fraction that is most accurate for *VSS-452*. As a result, our recommended DFAs maintain the rank ordering of probable top-5 DFAs compared to the ground truth, leading to a Spearman's $r$ of 0.95. This extremely high correspondence demonstrates that our DFA recommender is capable of identifying DFAs that are most likely to be accurate in a given chemical space.

**Interpreting TL model and recommender predictions.** We use virtual adversarial attack[18] as an approach to uncover TL model focus when predicting $|\Delta\Delta E_{H-L}[f]|$. During the attack, we obtain the virtual adversarial perturbation ($r_{vadv}$) on the inputs that maximizes the change of TL model output (i.e., before and after the perturbation), which represents the region of model focus. For example, the $r_{vadv}$ of the B3LYP TL model on Co(II)(SH$_2$)$_4$(SCN$^-$)$_2$ mostly concentrates on Co and S atoms, suggesting that the B3LYP TL model focuses mostly on the metal and first coordination sphere when predicting $|\Delta\Delta E_{H-L}|$ (Fig. 3a). By directly averaging over $r_{vadv}$ of all TMCs in the set-aside test set of *VSS-452* for a given DFA, we obtain its average TL model focus. Interestingly, we find that all TL models have much stronger focus on the metal-local environments compared to the untrained model, which is in agreement with our previous work[7,19] (Fig. 3b). This trend holds for all 48 DFAs despite the fact that spin-splitting energy itself can be sensitive to DFA choice and the associated HF exchange fraction.

Intuitively, one would expect the accuracy of a DFA for predicting spin splitting energy to depend on the ligand field strength of the ligands in the TMC[17,20]. This relationship, however, is challenging to disentangle when we have 48 candidate DFAs and only 152 test complexes, making all the statistics insignificant. To tease out the trends of the recommended DFAs with respect to the ligand field (measured by DLPNO-CCSD(T) $\Delta E_{H-L}$), we perform a control experiment using only a small pool of candidate DFAs that contain the best DFA at each semi-local functional (i.e., BLYP:50%, PBE:30%, SCAN:40%, M06-L:40%, and MN15-L:50%). For this experiment, we are able to maintain reasonable recommender performance (i.e., MAE = 2.5 kcal/mol) with these five select DFAs and still have a relatively large ratio between the number of test TMCs and candidate DFAs to make our statistical analysis sound. We further partition the

DLPNO-CCSD(T) $\Delta E_{H-L}$ from -100 kcal/mol to -10 kcal/mol equally into 9 ranges, thus quantifying the complexes from those containing the weakest to the strongest ligand field ligands. Among the five DFAs, each DFA has a ligand field strength range over which it performs the best (Fig. 4a). For example, M06-L:40% is mostly selected for strong ligand fields (DLPNO-CCSD(T) $\Delta E_{H-L}$ > -20 kcal/mol), while MN15-L:50% is recommended frequently for the weakest fields (DLPNO-CCSD(T) $\Delta E_{H-L}$ < -70 kcal/mol). This preference of selecting different DFAs at different ligand field strengths also explains the great accuracy of our recommender approach: At each specific range of DLPNO-CCSD(T) $\Delta E_{H-L}$, the recommender successfully avoids selecting DFAs that have a large MAE, which in turn will likely yield low errors in practical applications of the recommender (Fig. 4b).

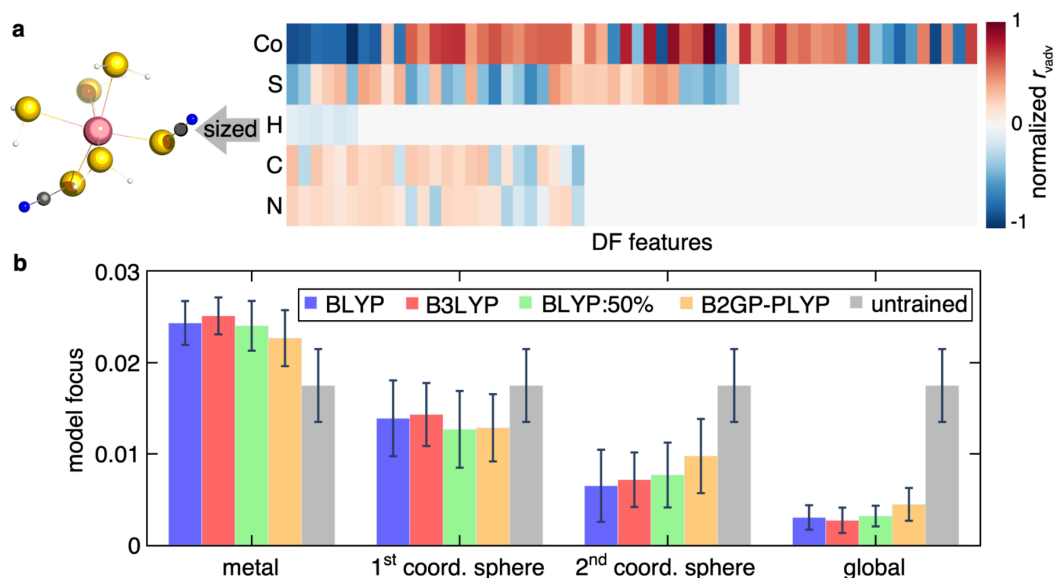

**Fig. 3 | Analysis of TL model focus using virtual adversarial attack. a**, The *cis* Co(II)(SH$_2$)$_4$(SCN)$_2$ molecule (left) where the sphere radius of each atom is proportional to its unsigned average of $r_{vadv}$ and its normalized $r_{vadv}$ (right) with the atom type shown at the start of each row. Only the average $r_{vadv}$ over each atom type is shown, since the differences are small within the same atom type due to the symmetry of this complex. The number of non-zero elements differs in $r_{vadv}$ since the basis set size varies for different atom types. All atoms are colored as follows: pink for Co, yellow for S, gray for C, blue for N, and white for H. **b**, Model focus decomposed to metal locality for select DFAs in BLYP family (blue for BLYP, red for B3LYP, green for BLYP:50%, and orange for B2GP-PLYP). Model focus of an untrained (i.e.,

randomly initialized) TL model is also shown as a comparison. Error bars represent the standard deviation across different TMCs in *VSS-452* set.

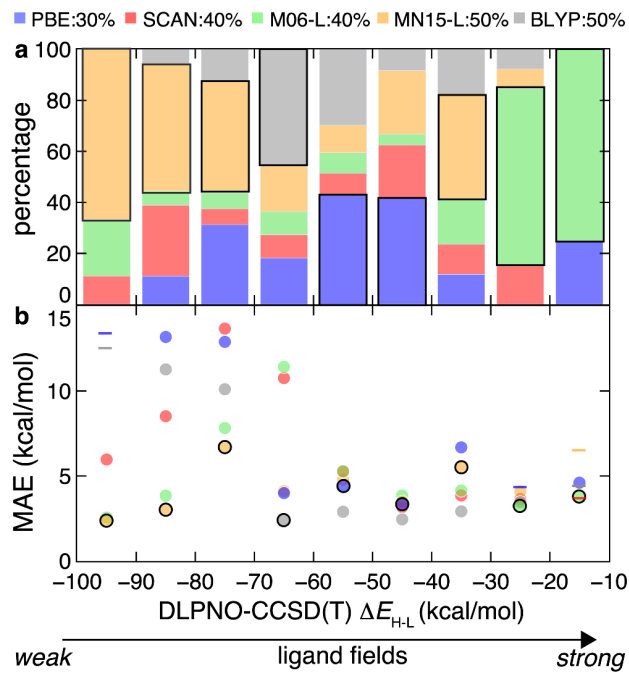

**Fig. 4 | Recommended DFAs by ligand field strength. a**, Stacked normalized histogram for the recommender-selected DFA by DLPNO-CCSD(T) $\Delta E_{\text{H-L}}$ with a bin width of 10 kcal/mol. **b**, $|\Delta\Delta E_{\text{H-L}}|$ MAE for the same DFAs (circles colored as in top legend) at different ranges of DLPNO-CCSD(T) $\Delta E_{\text{H-L}}$ grouped by the same set of bins in **a**. If a DFA is never selected in a range, it is shown with a horizontal bar instead of a circle. In both **a** and **b**, the DFA get most frequently selected in a range is outlined with a black solid outline. For ease of visualization, we show the recommender results on the set-aside test set of *VSS-452* with only five DFAs (blue for PBE:30%, red for SCAN:40%green for M06-L:40%, orange for MN15-L:50%, gray for BLYP:50%) as candidates.

**Transferability of the DFA recommender on diverse CSD complexes.** A more challenging test of the DFA recommender is on its application of chemically distinct *out-of-distribution* complexes. For this purpose, we construct *CSD-76*, a set of 76 TMCs randomly sampled from Cambridge Structural Database (CSD)[21] that contain diverse ligand chemistry, symmetry, and connectivity (Supplementary Fig. 9). Since all these complexes have been experimentally synthesized and crystallized, they test the DFA recommender on a realistic task for exploring transition metal chemical space. Without seeing any TMCs in *CSD-76*, the 48 TL models have

MAEs of predicting $|\Delta\Delta E_{H-L}[f]|$ that range from 3.1 to 6.6 kcal/mol, with a median of 4.5 kcal/mol (Fig. 5a). These MAEs are < 2 times those on the set-aside test set of *VSS-452*. While there is some accuracy degradation, the transferability improves significantly over chemical-composition-based representations that often yield > 5 times MAE on *out-of-distribution* CSD data[22]. We ascribe the great transferability of our TL models to the use of electron density as inputs, which is a more fundamental property and can, in principle, dictate all ground state properties of a system.

Using the predicted $|\Delta\Delta E_{H-L}[f]|$ from all 48 TL models, the recommender has a MAE of 3.0 kcal/mol, which still slightly outperforms the best TL model (3.1 kcal/mol from M06-2X, Fig. 5a). The recommender MAE on *CSD-76* is only 1.5 times that on *VSS-452* and is still within the threshold of accuracy required for transition metal chemical discovery compared to experimental uncertainties on measuring thermodynamic properties[16]. Despite the diverse and unseen chemistry present in *CSD-76*, the recommender still selects DFAs with < 3 kcal/mol error 60% of the time and < 5 kcal/mol error most of the time (82%), demonstrating its great transferability (Fig. 5b). These observations are particularly encouraging as the TL models' MAEs are more varied on *CSD-76* compared to *VSS-452* (Fig. 5a). Moreover, the TL model MAE rankings by dataset (e.g. *VSS-452* vs. *CSD-76*) are very different (Supplementary Figs. 4 and 10). For example, the DSD-PBEB95-D3BJ TL model gives the lowest MAE of 2.3 kcal/mol on *VSS-452* but a rather high MAE of 4.6 kcal/mol (i.e., the median of 48 TL models) on *CSD-76*. This highlights the robustness of the DFA recommender over the conventional TL approach, where the DFA recommender always give lower MAEs regardless the distributions and rankings of TL models built on different DFAs.

We next proceed to comparing the statistics of recommended DFAs for the *out-of-distribution CSD-76* and the set-aside test set of *VSS-452*. The recommender gives comparable ranking statistics of selected DFAs on *CSD-76*: 62% of the recommended DFAs are in the top 5 and 86% are in the top 10 (Supplementary Fig. 12). If we insisted on using the single "best" DFA benchmarked on the *VSS-452* set (i.e., DSD-BLYP-D3BJ) for exploring CSD chemical space, we would have a 5.90 kcal/mol MAE. Moreover, this functional choice is only the actual top-5-performing DFA 28% of the time over the *CSD-76* set. This observation, again, demonstrates the robustness and transferability of the recommender approach over both the conventional benchmark and TL approach on realistic chemical discovery.

Similar to the case of *VSS-452*, the DFA recommender is also able to identify the top-performing DFAs and correctly predict the relative likelihood of a DFA to be accurate for *CSD-76* (Fig. 5c). For example, it successfully identifies M06 as the most probable DFA to reside in the top-5, despite a slight overestimate of its likelihood (i.e., 64%) compared to the ground truth of 56%. In addition, the recommender maintains the high rank ordering (Spearman's $r = 0.90$) of probable top-5 DFAs relative to the ground truth on the *CSD-76* set, demonstrating its great transferability to unseen chemistry.

Due to the drastically different chemistry present in *VSS-452* and *CSD-76*, it is no surprise that the top-performing DFAs will vary for the two data sets (Supplementary Table 3, Figs. 3 and 10). MN15-L:50%, the most probable (45%) DFA residing in the top-5 choices for *VSS-452*, only has a 17% likelihood of being in top-5 for *CSD-76*. Meanwhile, M06, which is a top-5 DFA only 30% of the time for *VSS-452*, becomes the most probable DFA to reside in the top-5 with a probability of 56%. Although the recommender does not have this prior knowledge for the two data sets, it still captures the trend well and selects M06 much more often than

MN15L:50% for TMCs in *CSD-76* (Fig. 5c). The recommender also "intelligently" down-selects DFAs that only perform well on *VSS-452* (e.g., SCAN:40%, MN15-L:40%, M06-L:40%) and up-selects DFAs that would perform well on *CSD-76* (e.g., LRC-ωPBEh, SCAN0, PBE:20%). These observations suggest that our DFA recommender can be reliably applied to explore diverse transition metal chemical spaces with high accuracy.

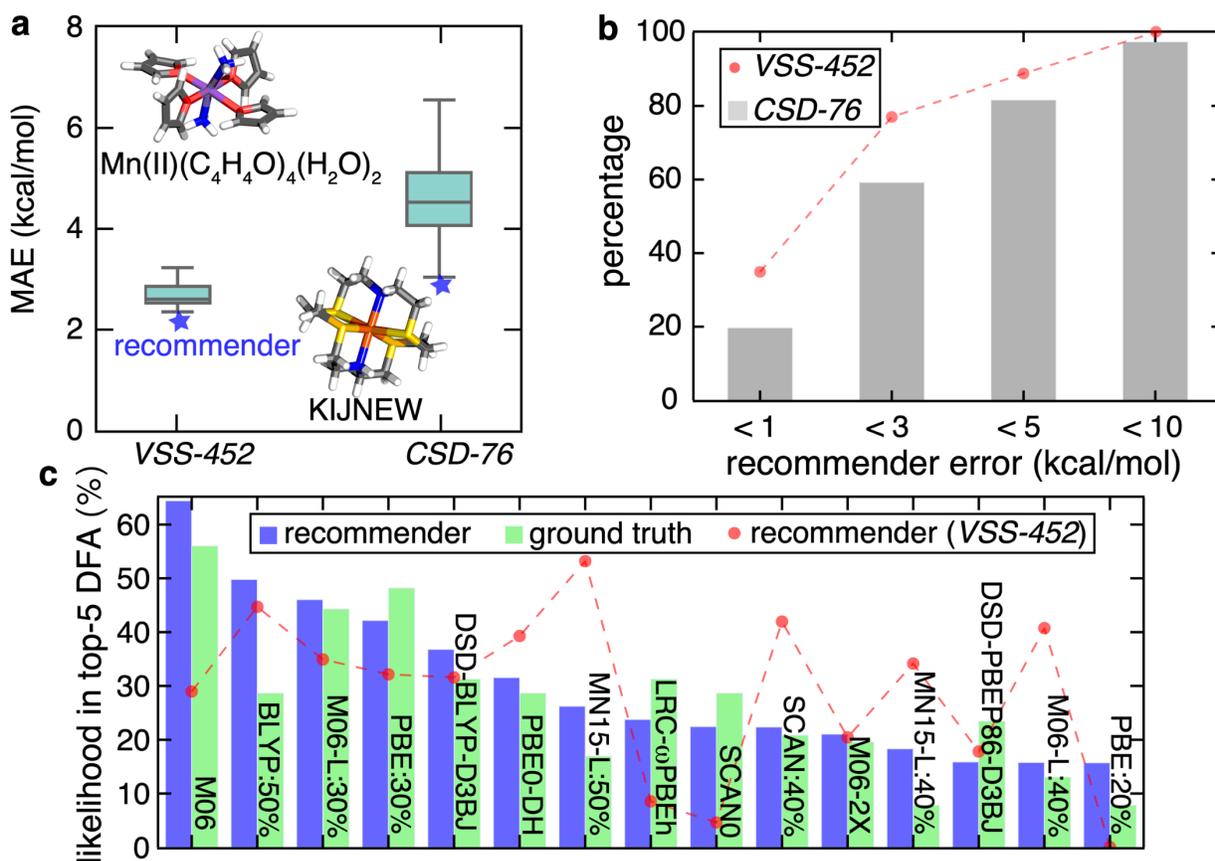

**Fig. 5 | Performance of TL models and the recommender on *CSD-76* set. a**, Box plot for MAE of 48 TL models for the prediction of $|\Delta\Delta E_{H-L}|$ for both the set-aside test set of *VSS-452* and the *CSD-76* set. The median of 48 TL models (solid line) and the recommender MAE (blue star) are also shown. For each data set, an example complex is shown: *trans* Mn(II)(C$_4$H$_4$O$_4$)$_4$(H$_2$O)$_2$ for *VSS-452* (left) and a hexadentate Fe complex (refcode: KIJNEW) for *CSD-76* (right). All atoms are colored as follows: purple for Mn, orange for Fe, yellow for S, gray for C, blue for N, red for O, and white for H. **b**, Percentage for the DFA recommender to have errors below certain thresholds (1, 3, 5, or 10 kcal/mol) for the set-aside test set of *VSS-452* (red circles) and *CSD-76* (gray bars). **c**, Percentage likelihood of a DFA residing in the top-5 choices suggested by ground truth (green) and our recommender approach (blue) for the *CSD-76* set, where DFAs are sorted in a descending order of the predicted likelihood of the recommender.

Percentage likelihood obtained by the recommender on the set-aside test set of *VSS-452* (red circles) is also shown as a comparison.

**Discussion**

DFT has become indispensable in both mechanistic study and in accelerated, automated chemical or materials discovery. Its accuracy, however, can highly depend on the choice of DFA. The single-DFA approach widely used in VHTS leads to bias in data acquisition, and expert knowledge and heuristics cannot be expected to be predictive of a single best DFA across large chemical spaces. In this work, we developed a general recommender approach to select DFAs in a system-specific manner. Distinct from traditional classification tasks where the label is certain, the "best" DFA can be expected to be ambiguous due to the similarities between candidate DFAs. We devise a "regress-then-classify" strategy to select DFAs with low error instead of forcing a model to directly classify the "best" DFA. By partitioning the electron density difference onto each atom within a system, we build Behler–Parrinello-type neural networks for transfer learning the differences between a DFA and the coupled cluster reference. The recommender then selects the DFA that gives the lowest predicted difference from the reference.

We demonstrated this recommender approach on evaluating the vertical spin splitting energy of open shell transition metal complexes. Trained only on 300 TMCs with common monodentate ligands, our recommender achieves an accuracy of 2.1 kcal/mol, outperforming both the conventional, single DFA and TL approach. This recommender also accurately captures the rank ordering (Spearman's $r$=0.96) of the likelihood of a DFA residing in the top-5 choices relative to the ground truth. When directly applied on experimentally synthesized complexes with diverse and unseen ligand chemistry and symmetry, the recommender maintains its stellar performance despite the fact that the top-performing DFAs for the CSD complexes are significantly different than the top-performing DFAs in the training data. The recommender still provides the accuracy needed for transition metal chemistry exploration (i.e., MAE=3.0 kcal/mol) and is able to select top-5 DFAs 62% of the time and capture the rank ordering (Spearman's $r$=0.90) of the likelihood of a DFA residing in the top-5 choices.

The recommender approach has two limitations in its current implementation. First, since it uses the B3LYP electron density as inputs, this recommender is not "zero cost", and its advantage is therefore greatest in transfer learning tasks where a property prediction with "beyond DFT" accuracy is needed. The use of ML-predicted density, semi-empirical densities, or guess densities (e.g., superposition of atomic potentials) can reduce the cost further. Second, a DFA may not be universally accurate across all properties for a system. Generalization of the current recommender may include redefining the loss function in the TL models to explicitly encode all relevant objectives.

In its present form, this recommender approach does not introduce additional computational cost when combined with existing DFT-based VHTS workflows that natively output an optimized geometry and electron density of a molecule. Therefore, it can be directly used in conjunction with traditional VHTS for improving the data quality from VHTS at no additional cost. In addition, our recommender approach is not restricted to predicting a single electronic energy of a molecule and thus can be generalized to more complex applications such as catalysis. Although we demonstrate the recommender to select from a pool of conventional DFAs, it is a general approach for method selection, including among semi-empirical theories, ML-derived DFAs, or wavefunction theories. We expect this recommender approach to be broadly useful in light of continuing advances in the methods available in the computational sciences.

## Methods

**Density fitting procedure.** In Kohn–Sham (KS) DFT, it is known that the ground state energy of any interacting system is captured by a universal functional of the electron density[23]. In practice, the electron density ($\rho$) is obtained from the occupied KS orbitals $\psi(r)$, expanded as a linear combination of the products of one-electron basis functions $\chi(r)$,

$$\rho(r) = \sum_i |\psi_i(r)|^2 = \sum_{\mu\nu} D_{\mu\nu} \chi_\mu(r) \chi_\nu(r) \quad (1)$$

where $D$ is the density matrix and $\mu$ and $\nu$ are indices for one-electron basis functions. The electron density in Eq. 1, however, is not expressed in an atom-centered basis and thus cannot be directly used as representation in neural networks. Thus, it is common to use density-fitting (DF) basis functions to rewrite the electron density as an expansion of atom-centered densities,

$$\rho(r) = \sum_A \sum_Q C_Q^A \phi_Q(r - r_A) = \sum_A \rho_A(r) \quad (2)$$

where $\phi_Q(r-r_A)$ is the $Q^{th}$ DF basis function for atom $A$[13]. However, $C_Q^A$ contains elements resulting from DF basis sets where the angular momentum is nonzero ($L \neq 0$) and is thus not rotationally invariant. To obtain a rotationally invariant representation, we calculated the power spectrum of $C_Q^A$ as the norm for each angular momentum $L$ in the DF basis set.

$$p_L^A = \sum_{Q \in L} \left\| C_Q^A \right\|^2 \quad (3)$$

Therefore, $p_L^A$ satisfies rotational, translational, and permutation symmetry and correspondingly can be used as a set of features into any neural network architectures (Fig. 1b). These features then represent the chemical environment of atom $A$. For this procedure, we employ only the density obtained from B3LYP, regardless of which functional is being studied in the TL models.

Here, we consider the vertical spin-splitting energy $\Delta E_{H-L}$ as our property of interest, which is the electronic energy difference between the high-spin (HS) and low-spin (LS) states of open-shell TMCs. We focus on the spin-splitting energy because it identifies the quantum mechanical ground state, which is an essential property of an open-shell system[4]. Because our target property involves two distinct spin states in open-shell systems, we decomposed the difference between the HS and LS electron densities for both the majority spin and minority spin separately (Fig. 1a).

$$\rho_{HS^\alpha}(r) - \rho_{LS^\alpha}(r) = \sum_A \sum_Q C_Q^{(A,\alpha)} \phi_Q(r - r_A) \quad (5)$$

$$\rho_{HS^\beta}(r) - \rho_{LS^\beta}(r) = \sum_A \sum_Q C_Q^{(A,\beta)} \phi_Q(r - r_A) \quad (6)$$

For an atom $A$, we obtained and concatenated the power spectra of $C_Q^{(A,\alpha)}$ and $C_Q^{(A,\beta)}$ to obtain its features using Eq. 3. We used the def2-universal-jkfit[24] as our DF basis set throughout this work. The number of DF features for different atoms can vary due to the differences in the auxiliary basis functions used by atoms. Here, we zero-padded the DF features for all atoms to the maximum dimension of 58 for each atom density, which is the size of the DF basis set for the transition metal atom (i.e., Cr, Mn, Fe, or Co) in a TMC.

**Behler–Parrinello-type neural networks for transfer learning.** We built Behler–Parrinello-type neural networks using the DF representation of the TMCs in this work[25]. These fully-connected neural networks used the DF representation of each atom as inputs,

$$X_A^l = \sigma(W_{A \in g}^l X_A^{l-1}) \quad (7)$$

where $X_A^l$ is the representation of atom $A$ at layer $l$, $W_{A \in g}^l$ is the $l^{\text{th}}$-layer weights for the network of elements in group $g$, and $\sigma$ is the activation function. Specifically, $X_A^0$ is the set of concatenated DF features of atom $A$ (see *density fitting procedure*). The last layer of the network outputs, $X_A^n$, are summed for each chemical element ($e$),

$$X_e^n = \sum_{A \in e} X_A^n \qquad (8)$$

These $X_e^n$ of different elements are then concatenated and passed to a fully-connected neural network to obtain the final output (Fig. 1c).

Our model has three main differences from the original Behler–Parrinello neural network. First, we replace the symmetry functions that describe the local geometric environment of an atom therein by the DF representation, which is derived from the electron density and is thus a more transferable representation. Second, we use the same local network for chemical elements that are in the same group of the periodic table (e.g., O and S) to promote inter-row learning[26]. Lastly, we keep the latent vector $X_e^n$ for each element and use a neural network to obtain the final output because our final target is not an single electronic energy of the ground state.

We adopted TL strategies and chose our target to be the absolute difference of vertical spin-splitting energies between the result from each DFA ($f$) and a reference calculation ($|\Delta\Delta E_{\text{H–L}}[f]|$). For each fully-connected neural network, we used three hidden layers and 96 neurons per layer. The shifted softplus activation function, $\sigma(x) = \text{softplus}(x) - \log(2)$, is used throughout.

**Recommender.** We constructed separate TL models for each DFA ($f$) to predict $|\Delta\Delta E_{\text{H–L}}[f]|$ from a pre-selected pool of DFAs ($F$). For a given system, we recommend the DFA, $f_{\text{rec}}$, that yields the lowest predicted $|\Delta\Delta E_{\text{H–L}}[f]|$,

$$f_{\text{rec}} = \text{argmin}_{f \in F} |\Delta\Delta E_{\text{H-L}}[f]| \quad (9)$$

When we evaluate the practical performance of the DFA recommender, we focus on the absolute error introduced by using $f_{\text{rec}}$ relative to the reference method (i.e., $|\Delta\Delta E_{\text{H-L}}[f]|$) and the actual ranking of $f_{\text{rec}}$ among the pool of DFAs.

**Data set construction.** Mononuclear octahedral TMCs with Cr, Mn, Fe, and Co in oxidation states II and III were studied in their HS and LS states: quintet and singlet for $d^6$ Co(III)/Fe(II) and $d^4$ Mn(III)/Cr(II); sextet and doublet for $d^5$ Fe(III)/ Mn(II), and quartet and doublet for $d^3$ Cr(III) and $d^7$ Co(II). For *VSS-452*, we used 20 monodentate ligands from both the spectrochemical series and common organic ligands to obtain properties of complexes with ligand fields ranging from weak to strong (Supplementary Fig. 2). We allowed up to two unique ligands in a TMC and did not pose any constraints on ligand symmetry. Together with eight metal–oxidation state combination and 20 ligands, this rule of assembling TMCs leads to a hypothetical space of 24,480 TMCs (8×20=160 homoleptic and 8×20×19×8=24,320 heteroleptic). We used *k*-medoids sampling to obtain 750 TMCs in this space as our starting data set. To test the transferability and practical usefulness of our recommender, we collected 100 experimentally synthesized TMCs with diverse ligand chemistry and connectivity from CSD as the starting point for *CSD-76*.

**DFT geometry optimization.** Since we are interested in vertical spin splitting, only one structure needs to be geometry optimized. In this case, we chose to optimize only the HS state. For each HS complex, a DFT geometry optimization with the B3LYP[27] global hybrid functional was carried out using a developer version of graphical processing unit (GPU)-accelerated electronic structure code TeraChem[28]. The LANL2DZ effective core potential[29] basis set was used for metals and the 6-31G* basis[24] for all other atoms. In all DFT geometry optimizations, level shifting[30] of 0.25 Ha on all virtual orbitals was employed. Initial geometries were assembled by molSimplify[31] for *VSS-452* and were adopted from the crystal structure of CSD for *CSD-76*. These geometries were optimized using the L-BFGS algorithm in translation rotation internal coordinates (TRIC)[32] to the default tolerances of $4.5 \times 10^{-4}$ hartree/bohr for the maximum gradient and $10^{-6}$ hartree for the energy change between steps. Because all HS TMCs are open-shell, the unrestricted formalism was used for all geometry optimizations.

Geometry checks were applied to eliminate optimized structures that deviated from the expected octahedral shape following previously established metrics without modification[33]. Open-shell structures were also removed from the data set following established protocols if the expectation value of the $S^2$ operator deviated from its expected value[33] of $S(S + 1)$ by $>1$ $\mu_B^2$. After these two filtering steps, we converged 452 HS TMCs for *VSS-452* and 76 HS TMCs for *CSD-76* with good octahedral geometries and electronic structures (Supplementary Table 4).

**Single-point energy calculation.** We followed our established protocol for the computation of HS and LS electronic energies with multiple DFAs for the optimized TMCs using a developer version of Psi4 1.4[34]. In this workflow, the converged wavefunction obtained from the B3LYP geometry optimization was used as the initial guess for the single-point energy calculations with other DFAs, thus maximizing the correspondence of the converged electronic state among all DFAs and also reducing the computational cost.

The range of 23 DFAs used in the development of the protocol[7] were chosen to be evenly distributed among the rungs of "Jacob's ladder"[35] (Supplementary Table 1). Practically, it has been observed that there is a nearly linear change of chemical properties (e.g., spin splitting) computed with a DFA at different fractions of HF exchange[20]. Therefore, we sampled the HF exchange from 10% to 50% with an interval of 10% on five selected semi-local functionals (i.e., BLYP, PBE, SCAN, M06-L, and MN15-L). This procedure results in 25 additional DFAs (Supplementary Table 2). Combined with the original 23 DFAs, we have a final pool of 48 DFAs in total.

CCSD(T) has been treated as the "gold standard" for quantum chemistry and is frequently used as benchmark for DFT[17]. Here, we used DLPNO-CCSD(T) (with T0 perturbative triple correction[15]), which is a proxy for canonical CCSD(T), as our reference method due to the sufficient accuracy of DLPNO-CCSD(T) on TMCs and the high computational cost of canonical CCSD(T) for a large data set[15]. In addition, we expect our DFA recommender approach to be general and have similar accuracy if reference data is derived from higher-level theory (e.g., phaseless auxiliary field quantum Monte-Carlo[36]) or experiments in the future. Both DFT and DLPNO-CCSD(T) single-point energies for all non-singlet states were calculated with an unrestricted formalism and for singlet states with a restricted formalism. All single-point energy calculations were performed with a balanced polarized triple-zeta basis set def2-TZVP[24].

**Train/test partition and model training.** We randomly partitioned *VSS-452*, with 300 points (66%) as the training set and 152 (34%) points as the set-aside test set. For all TL models, the hyperparameters were selected using HyperOpt[37] with 200 evaluations, with 60 points of the training set used as the validation data. All TL models were built with PyTorch[38]. All models were trained with the Adam optimizer up to 2000 epochs, using dropout and early stopping to avoid over-fitting. We treated *CSD-76* as the out-of-distribution test set, and thus no points in *CSD-76* was used during the whole model training procedure.

Supplementary Information of "A transferable recommender approach for selecting the best density functional approximations in chemical discovery"

Chenru Duan[1,2], Aditya Nandy[1,2], Ralf Meyer[1], Naveen Arunachalam[1], and Heather J. Kulik[1,2]

[1]Department of Chemical Engineering, Massachusetts Institute of Technology, Cambridge, MA 02139

[2]Department of Chemistry, Massachusetts Institute of Technology, Cambridge, MA 02139

**Abbreviation**

The following is the list of abbreviation utilized in the main paper.

- DFA: Density functional approximation
- MAE: Mean absolute error
- VSS: vertical spin splitting
- CSD: Cambridge Structural Database
- HF: Hartree-Fock
- LRC: Long-range correction
- RS: Range-separated
- GGA: General gradient approximation

**Supplementary Figures**

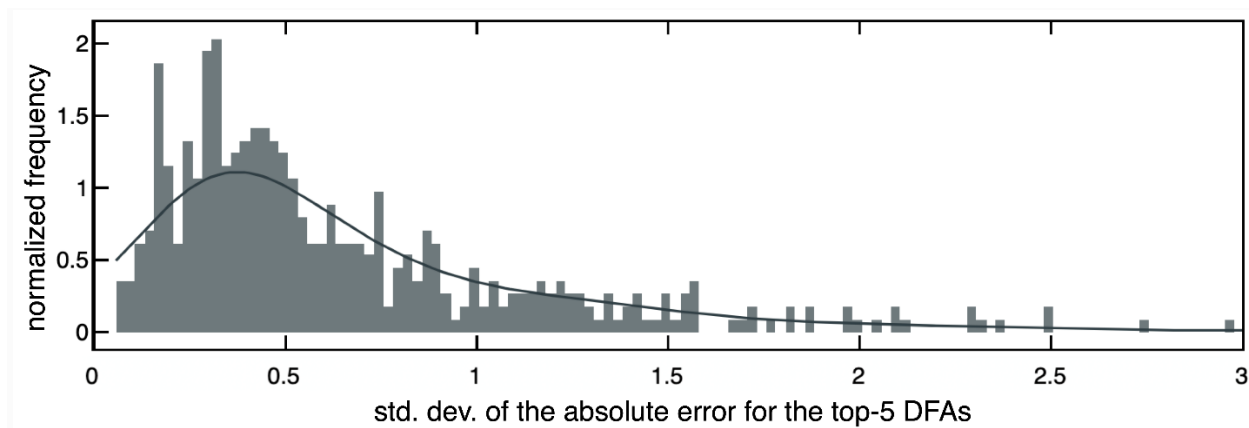

**Supplementary Figure 1. Standard deviation of the absolute error for the top-5 DFAs.** Normalized distributions of standard deviation (std. dev.) of $|\Delta\Delta E_{H-L}|$ for the top-5 DFAs. A kernel density estimate (black) is also shown. It can be observed that the differences of $|\Delta\Delta E_{H-L}|$ for the top-5 DFAs are mostly small (i.e., to 1.0 kcal/mol) for TMCs in *VSS-452* set.

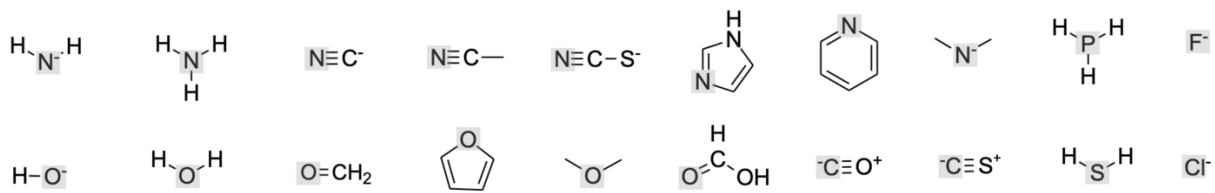

**Supplementary Figure 2. 20 ligands used in *VSS-452*.** The 20 small monodentate ligands resembling those from the spectrochemical series that are used for *VSS-452* set. The coordinating atom is shaded in gray for each ligand.

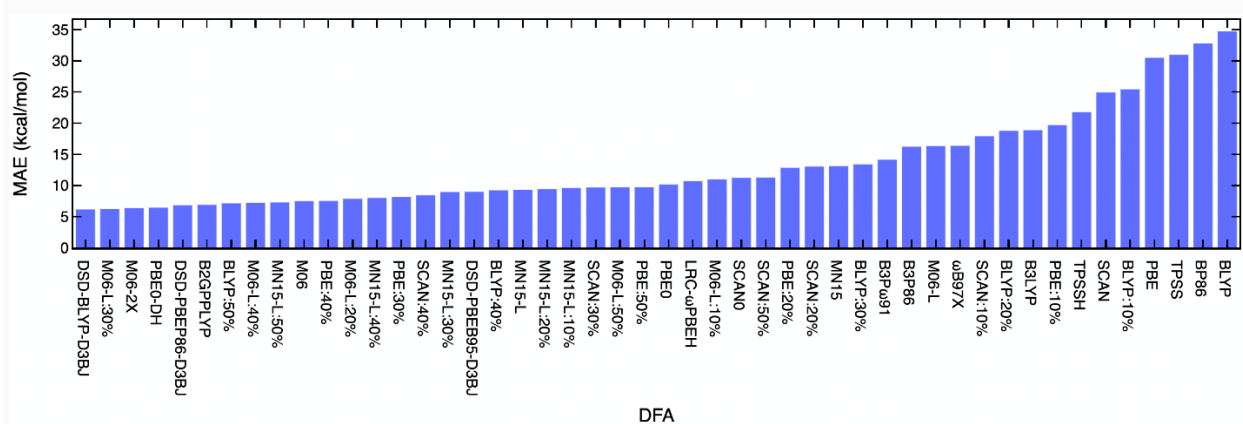

**Supplementary Figure 3. MAEs derived from the 48 DFAs compared to DLPNO-CCSD(T).** MAE of $|\Delta\Delta E_{H-L}|$ for the 48 DFAs of the *VSS-452* set. The DFAs are sorted in an ascending order of their DFA-derived MAEs.

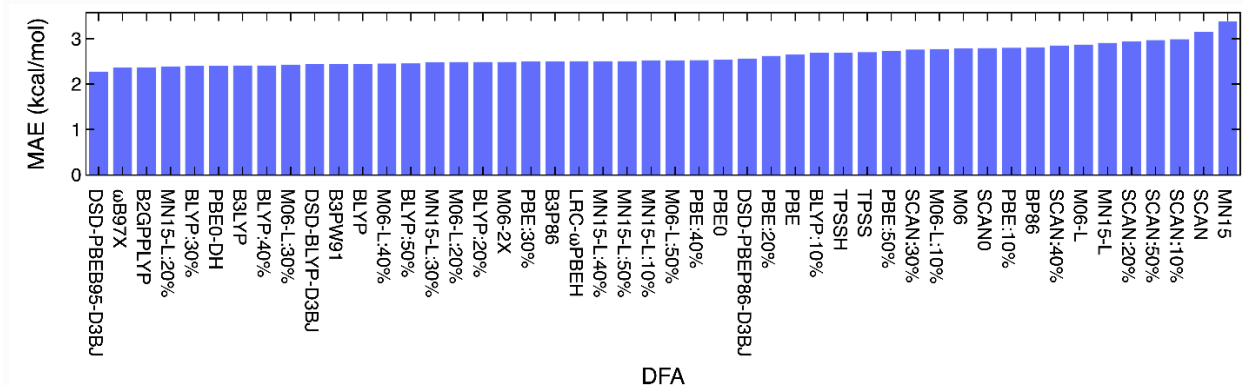

**Supplementary Figure 4. MAEs derived from the 48 TL models compared to DLPNO-CCSD(T).** MAE of $|\Delta\Delta E_{H-L}|$ for the 48 TL models of the set-aside test set of *VSS-452*. The DFAs are sorted in an ascending order of their TL MAEs.

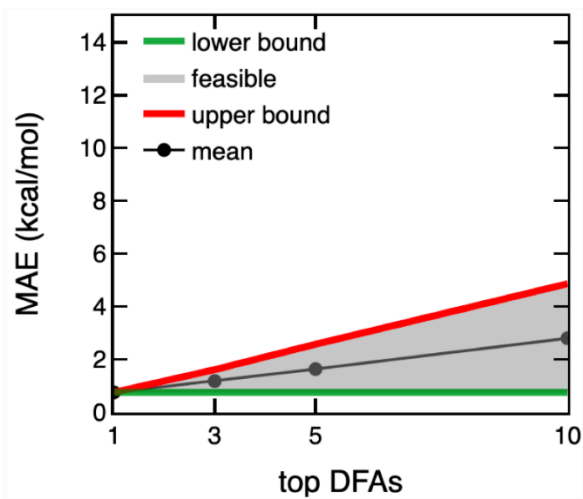

**Supplementary Figure 5. Theoretical bound of recommender approach with 48 DFAs.** MAE of $|\Delta\Delta E_{H-L}|$ constrained by only selecting a random top-n DFA (black dots and solid line) among the 48 DFAs. The lower bound (green line), upper bound (red line), and the feasible area (gray shaded) are also shown when the DFA selection is constrained in top-n choices.

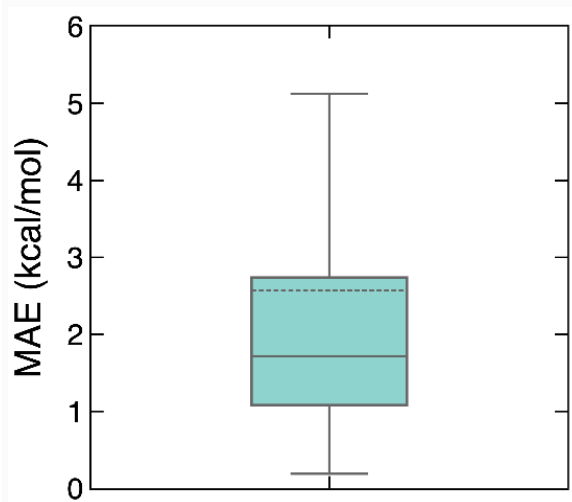

**Supplementary Figure 6. Box plot for the performance of $5^{th}$-best DFA.** MAE for the $5^{th}$-best DFA of $|\Delta\Delta E_{H-L}|$ with a box indicating their median (solid line) and mean (dashed line) for the *VSS-452* set.

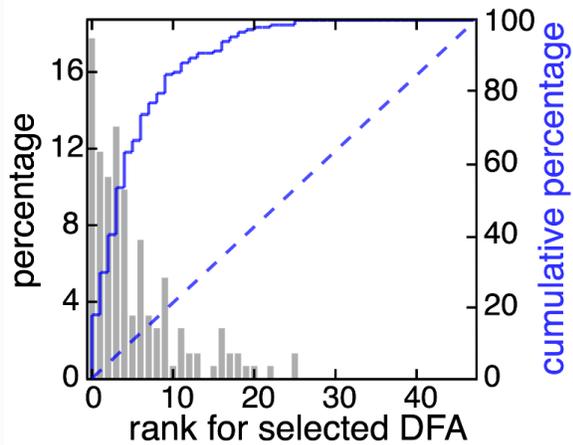

**Supplementary Figure 7. Recommender rank-ordering performance with 48 DFAs.** Normalized distribution of the rank for selected DFA using our recommender approach, with the cumulative percentage (blue solid line) shown according to the axis on the right. The cumulative curve for a random guess (blue dashed line) is also shown.

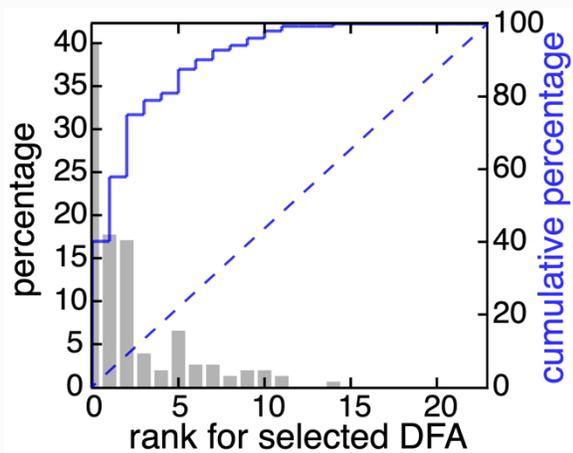

**Supplementary Figure 8. Recommender rank-ordering performance with 23 DFAs from Duan *et al.*[1].** Normalized distribution of the rank for selected DFA using our recommender approach, with the cumulative percentage (blue solid line) shown according to the axis on the right. The cumulative curve for a random guess (blue dashed line) is also shown.

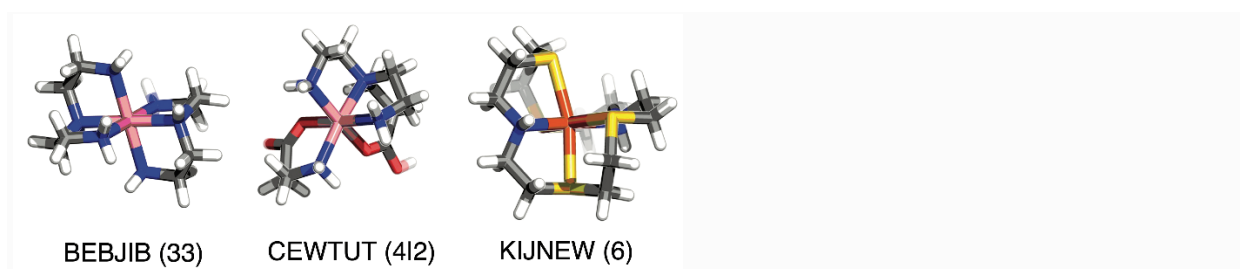

**Supplementary Figure 9. Example TMCs in CSD-76.** A Co complex with two tridentate ligands (refcode: BEBJIB, left), a Co complex with one tetradentate and one bidentate ligand (refcode: CEWTUT, middle), and a Fe complex with a hexadentate ligand (refcode: KIJNEW, right). All atoms are colored as follows: pink for Co, orange for Fe, yellow for S, gray for C, blue for N, red for O, and white for H.

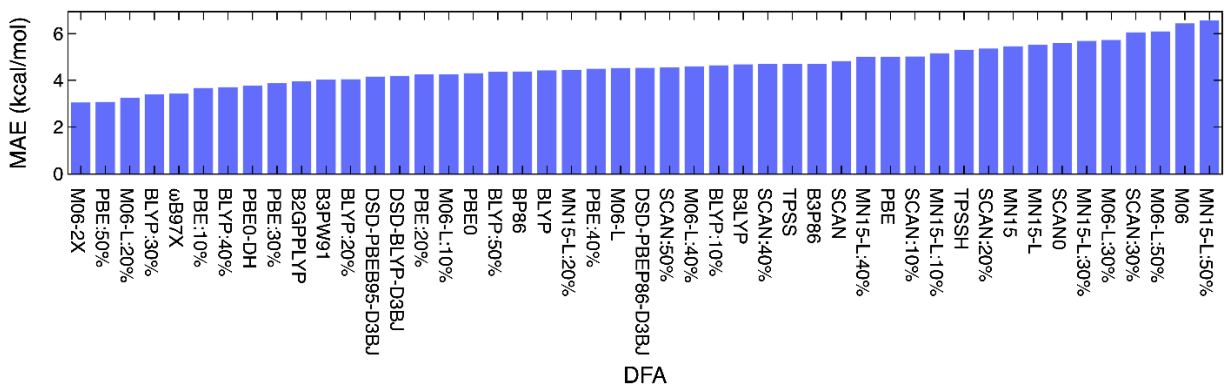

**Supplementary Figure 10. MAEs derived from the 48 TL models compared to DLPNO-CCSD(T).** MAE of $|\Delta\Delta E_{H-L}|$ for the 48 TL models of the *CSD-76* set. The DFAs are sorted in an ascending order of their TL MAEs.

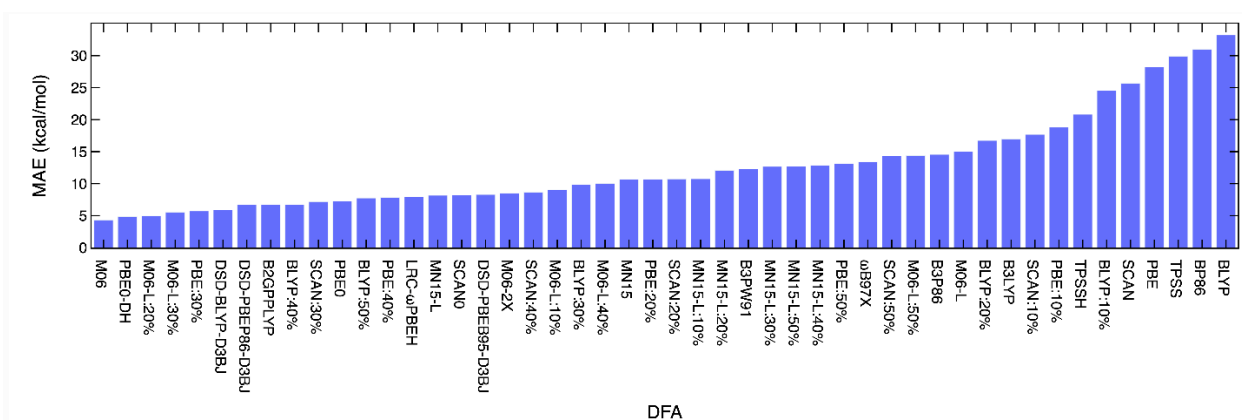

**Supplementary Figure 11. MAEs derived from the 48 DFAs compared to DLPNO-CCSD(T).** MAE of $|\Delta\Delta E_{H-L}|$ for the 48 DFAs of the *CSD-76* set. The DFAs are sorted in an ascending order of their DFA-derived MAEs.

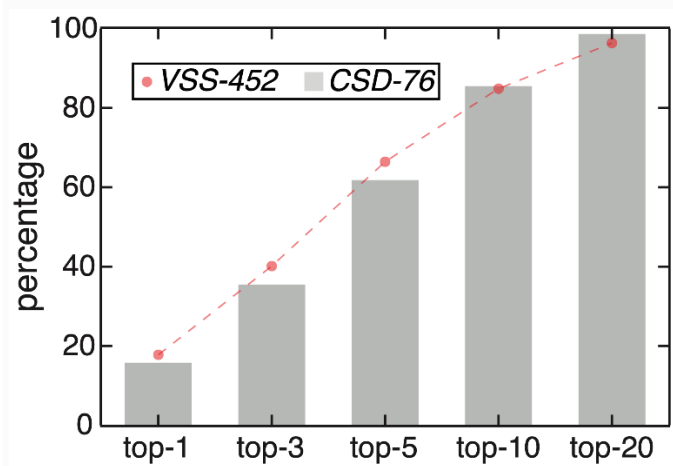

**Supplementary Figure 12. Comparison of ranking statistics for recommended DFAs of set-aside test *VSS-452* and *CSD-76*.** Percentage for the DFA recommender to select top-n DFAs for the set-aside test set of *VSS-452* (red circles) and *out-of-distribution CSD-76* (gray bars).

# Supplementary Tables

**Table 1.** Summary of 23 DFAs in the original work of Duan *et al.*[1], including their rungs on "Jacob's ladder" of DFT, HF exchange fraction, LRC range-separation parameter (bohr$^{-1}$), MP2 correlation fraction, and whether empirical (i.e., D3) dispersion correction is included.

| DFA | type | exchange type | HF exchange percentage | LRC RS parameter (bohr$^{-1}$) | MP2 correlation | D3 dispersion |
|---|---|---|---|---|---|---|
| BP86[2-3] | GGA | GGA | -- | -- | -- | no |
| BLYP[4-5] | GGA | GGA | -- | -- | -- | no |
| PBE[6] | GGA | GGA | -- | -- | -- | no |
| TPSS[7] | meta-GGA | meta-GGA | -- | -- | -- | no |
| SCAN[8] | meta-GGA | meta-GGA | -- | -- | -- | no |
| M06-L[9] | meta-GGA | meta-GGA | -- | -- | -- | no |
| MN15-L[10] | meta-GGA | meta-GGA | -- | -- | -- | no |
| B3LYP[11-13] | GGA hybrid | GGA | 0.200 | -- | -- | no |
| B3P86[2, 11] | GGA hybrid | GGA | 0.200 | -- | -- | no |
| B3PW91[11, 14] | GGA hybrid | GGA | 0.200 | -- | -- | no |
| PBE0[15] | GGA hybrid | GGA | 0.250 | -- | -- | no |
| ωB97X[16] | RS hybrid | GGA | 0.158 | 0.300 | -- | no |
| LRC-ωPBEh[17] | RS hybrid | GGA | 0.200 | 0.200 | -- | no |
| TPSSh[7] | meta-GGA hybrid | meta-GGA | 0.100 | -- | -- | no |
| SCAN0[18] | meta-GGA hybrid | meta-GGA | 0.250 | -- | -- | no |
| M06[19] | meta-GGA hybrid | meta-GGA | 0.270 | -- | -- | no |
| M06-2X[19] | meta-GGA hybrid | meta-GGA | 0.540 | -- | -- | no |
| MN15[20] | meta-GGA hybrid | meta-GGA | 0.440 | -- | -- | no |
| B2GP-PLYP[21] | double hybrid | GGA | 0.650 | -- | 0.360 | no |
| PBE0-DH[22] | double hybrid | GGA | 0.500 | -- | 0.125 | no |
| DSD-BLYP-D3BJ[23] | double hybrid | GGA | 0.710 | -- | 1.000 | yes |
| DSD-PBEB95-D3BJ[23] | double hybrid | GGA | 0.660 | -- | 1.000 | yes |
| DSD-PBEP6-D3BJ[23] | double hybrid | GGA | 0.690 | -- | 1.000 | yes |

**Table 2.** Summary of the additional 25 DFAs compared to Duan et al.[1], including their rungs on "Jacob's ladder" of DFT, Hartree–Fock (HF) exchange fraction, long-range correction (LRC) range-separation parameter (bohr$^{-1}$), MP2 correlation fraction, whether empirical (i.e., D3) dispersion correction is included.

| DFA | type | exchange type | HF exchange percentage | LRC RS parameter (bohr$^{-1}$) | MP2 correlation | D3 dispersion |
|---|---|---|---|---|---|---|
| BLYP:10% | GGA hybrid | GGA | 0.100 | -- | -- | no |
| BLYP:20% | GGA hybrid | GGA | 0.200 | -- | -- | no |
| BLYP:30% | GGA hybrid | GGA | 0.300 | -- | -- | no |
| BLYP:40% | GGA hybrid | GGA | 0.400 | -- | -- | no |
| BLYP:50% | GGA hybrid | GGA | 0.500 | -- | -- | no |
| PBE:10% | GGA hybrid | GGA | 0.100 | -- | -- | no |
| PBE:20% | GGA hybrid | GGA | 0.200 | -- | -- | no |
| PBE:30% | GGA hybrid | GGA | 0.300 | -- | -- | no |
| PBE:40% | GGA hybrid | GGA | 0.400 | -- | -- | no |
| PBE:50% | GGA hybrid | GGA | 0.500 | -- | -- | no |
| SCAN:10% | meta-GGA hybrid | meta-GGA | 0.100 | -- | -- | no |
| SCAN:20% | meta-GGA hybrid | meta-GGA | 0.200 | -- | -- | no |
| SCAN:30% | meta-GGA hybrid | meta-GGA | 0.300 | -- | -- | no |
| SCAN:40% | meta-GGA hybrid | meta-GGA | 0.400 | -- | -- | no |
| SCAN:50% | meta-GGA hybrid | meta-GGA | 0.500 | -- | -- | no |
| M06-L:10% | meta-GGA hybrid | meta-GGA | 0.100 | -- | -- | no |
| M06-L:20% | meta-GGA hybrid | meta-GGA | 0.200 | -- | -- | no |
| M06-L:30% | meta-GGA hybrid | meta-GGA | 0.300 | -- | -- | no |
| M06-L:40% | meta-GGA hybrid | meta-GGA | 0.400 | -- | -- | no |
| M06-L:50% | meta-GGA hybrid | meta-GGA | 0.500 | -- | -- | no |
| MN15-L:10% | meta-GGA hybrid | meta-GGA | 0.100 | -- | -- | no |
| MN15-L:20% | meta-GGA hybrid | meta-GGA | 0.200 | -- | -- | no |
| MN15-L:30% | meta-GGA hybrid | meta-GGA | 0.300 | -- | -- | no |
| MN15-L:40% | meta-GGA hybrid | meta-GGA | 0.400 | -- | -- | no |
| MN15-L:50% | meta-GGA hybrid | meta-GGA | 0.500 | -- | -- | no |

**Table 3.** Ranking of DFA-derived MAEs of 48 DFAs on *VSS-452* and *CSD-76*.

| DFA | MAE ranking in *VSS-452* | MAE ranking in *CSD-76* |
|---|---|---|
| DSD-BLYP-D3BJ | 1 | 6 |
| M06-L:30 | 2 | 4 |
| M06-2X | 3 | 18 |
| PBE0-DH | 4 | 2 |
| DSD-PBEP86-D3BJ | 5 | 7 |
| B2GP-PLYP | 6 | 8 |
| BLYP:50% | 7 | 12 |
| M06-L:40% | 8 | 22 |
| MN15-L:50% | 9 | 30 |
| M06 | 10 | 1 |
| PBE:40% | 11 | 13 |
| M06-L:20% | 12 | 3 |
| MN15-L:40% | 13 | 31 |
| PBE:30% | 14 | 5 |
| SCAN:40% | 15 | 19 |
| MN15-L:30% | 16 | 29 |
| DSD-PBEB95-D3BJ | 17 | 17 |
| BLYP:40% | 18 | 9 |
| MN15-L | 19 | 15 |
| MN15-L:20% | 20 | 27 |
| MN15-L:10% | 21 | 26 |
| SCAN:30% | 22 | 10 |
| M06-L:50% | 23 | 35 |
| PBE:50% | 24 | 32 |
| PBE0 | 25 | 11 |
| LRC-wPBEh | 26 | 14 |
| M06-L:10% | 27 | 20 |
| SCAN0 | 28 | 16 |
| SCAN:50% | 29 | 34 |
| PBE:20% | 30 | 24 |
| SCAN:20% | 31 | 25 |
| MN15 | 32 | 23 |
| BLYP:30% | 33 | 21 |
| B3PW91 | 34 | 28 |
| B3P86 | 35 | 36 |
| M06-L | 36 | 37 |
| w97X | 37 | 33 |
| SCAN:10% | 38 | 40 |
| BLYP:20% | 39 | 38 |
| B3LYP | 40 | 39 |
| PBE:10% | 41 | 41 |
| TPSSh | 42 | 42 |
| SCAN | 43 | 44 |
| BLYP:10% | 44 | 43 |
| PBE | 45 | 45 |
| TPSS | 46 | 46 |
| BP86 | 47 | 47 |
| BLYP | 48 | 48 |

**Table 4.** Summary of data filtering statistics for *VSS-452* and *CSD-76*.

|  | attempted | converged with good geometry and electronic structure |
|---|---|---|
| *VSS-452* | 750 | 452 |
| *CSD-76* | 100 | 76 |